\documentclass{PoS}
\usepackage[percent]{overpic}
\title{Study of the phase diagram of strongly interacting matter in
the NA61/SHINE experiment}

\ShortTitle{Strong interactions in NA61/SHINE}

\author{\speaker{Maja Ma\'{c}kowiak-Paw{\l}owska} for the NA61/SHINE Collaboration\\
        Faculty of Physics, Warsaw University of Technology\\
        E-mail: \email{maja.pawlowska@pw.edu.pl}}

\abstract{NA61/SHINE (SPS Heavy Ion and Neutrino Experiment) is a fixed target experiment located at the CERN SPS. Its strong interactions program is devoted to study properties of the phase diagram of strongly interacting matter. For this goal the two-dimensional scan is performed by measurements of hadron production properties as a function of collision energy (13A - 158A GeV/$c$) and system size (p+p, p+Pb, Be+Be, Ar+Sc, Xe+La, Pb+Pb). This contribution presents new results on the onset of deconfinement - the transition between the state of hadronic matter and the quark-gluon plasma. Also, new results on fluctuations and correlations devoted to the search for the critical point of strongly interacting matter will be presented. Obtained results are compared with the available data from other experiments and from various theoretical models.}

\FullConference{%
  *** Particles and Nuclei International Conference - PANIC2021 ***\\
  *** 5 - 10 September, 2021 ***\\
  *** Online ***
}

\begin{document}

\section{Introduction}
NA61/SHINE~\cite{Abgrall:2014xwa} at the CERN Super Proton Synchrotron (SPS) is a fixed-target experiment pursuing a rich physics program including measurements for strong interactions, neutrino, and cosmic ray physics.

\begin{figure}[t]
			\centering	
			\includegraphics[width=0.45\textwidth]{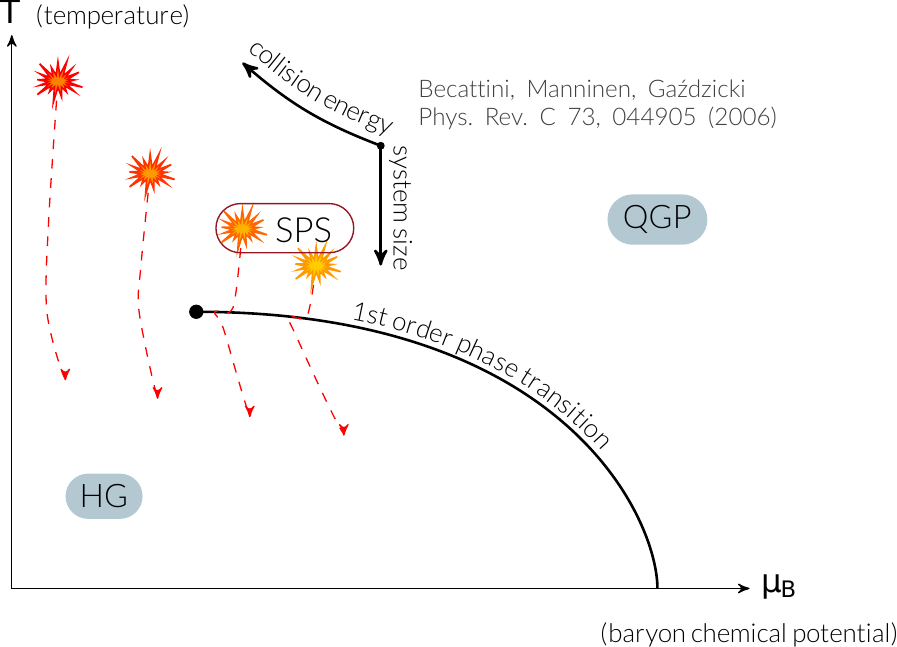}
			\includegraphics[width=0.45\textwidth]{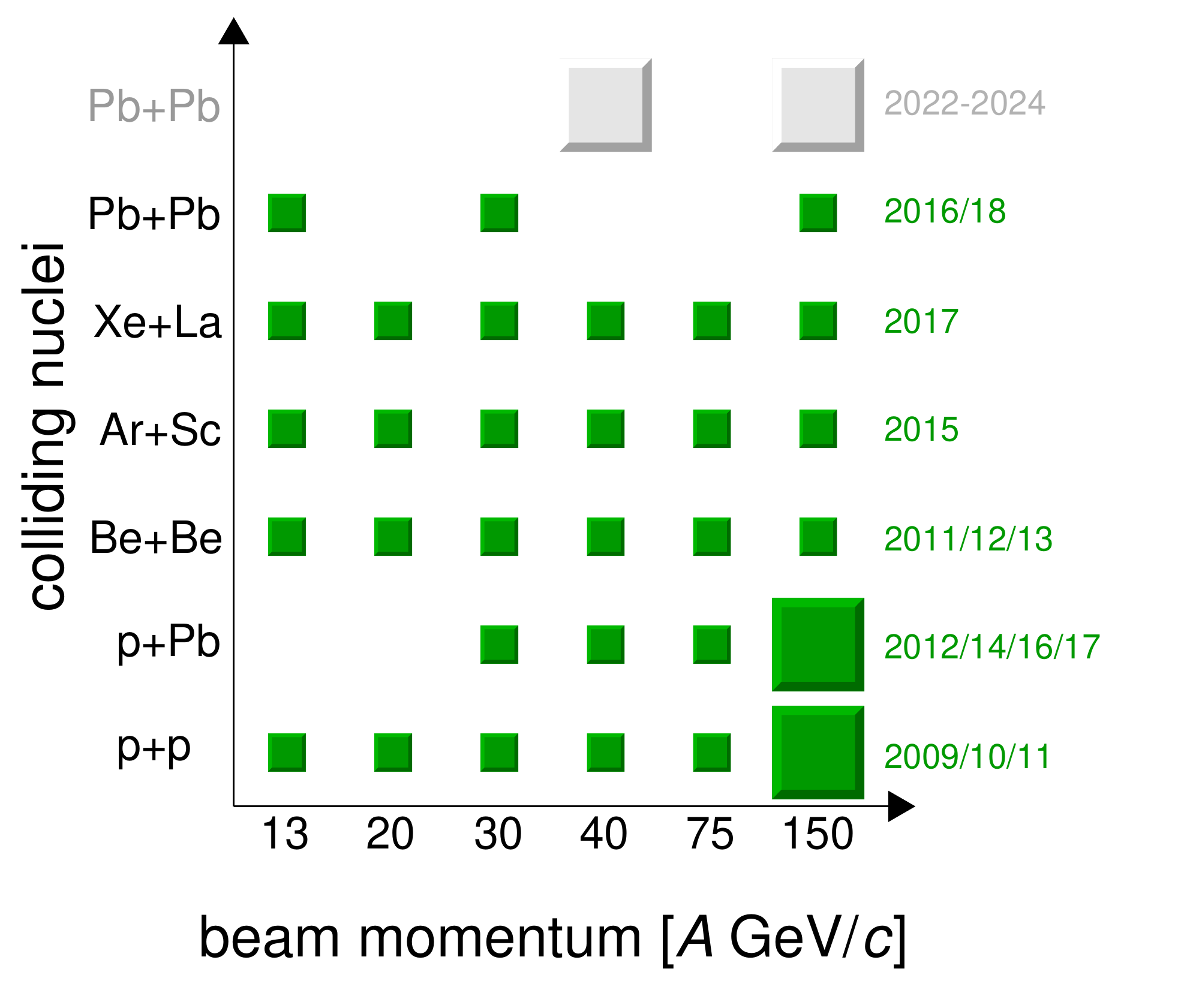}			
			\vspace{-0.15in}
			\caption{Illustration of phase-diagram of strongly interacting matter and NA61/SHINE system size and energy scan}
			\label{fig:program}
	\end{figure}

Among the aims of the strong interactions program is to search for the critical point (CP) and study of the onset of deconfinement (OD) of strongly interacting matter. NA61/SHINE is the first experiment which explores the phase diagram of strongly-interacting matter by performing a two-dimensional scan, in beam momentum (13$A$ -- 150/158$A$ GeV/$c$)  and size  of colliding system (p+p, p+Pb, Be+Be, Ar+Sc, Xe+La, Pb+Pb). This is exactly the energy range where indications of OD were reported~\cite{NA491}. Such a scan allows for a rare possibility for measurements of strong interactions at different conditions with similar detector set-up. The illustration of the scan as well as list of gathered and planed system/energies are shown in Fig.~\ref{fig:program} along with the illustration of the phase-diagram of strongly interacting matter.

This contribution discusses new results on the OD studies as well as on the search for the CP.
	
\section{Study of the onset of deconfinement}

The Statistical Model of the Early Stage (SMES)~\cite{SMES} predicts several signatures of the $1^{st}$ order phase transition from hadrons to quarks and gluons. In the transition region, constant temperature and pressure in the mixed-phase and an increase of the number of internal degrees of freedom is expected. 

\begin{figure}[t]
			\begin{minipage}{0.65\linewidth}
			\includegraphics[width=\textwidth]{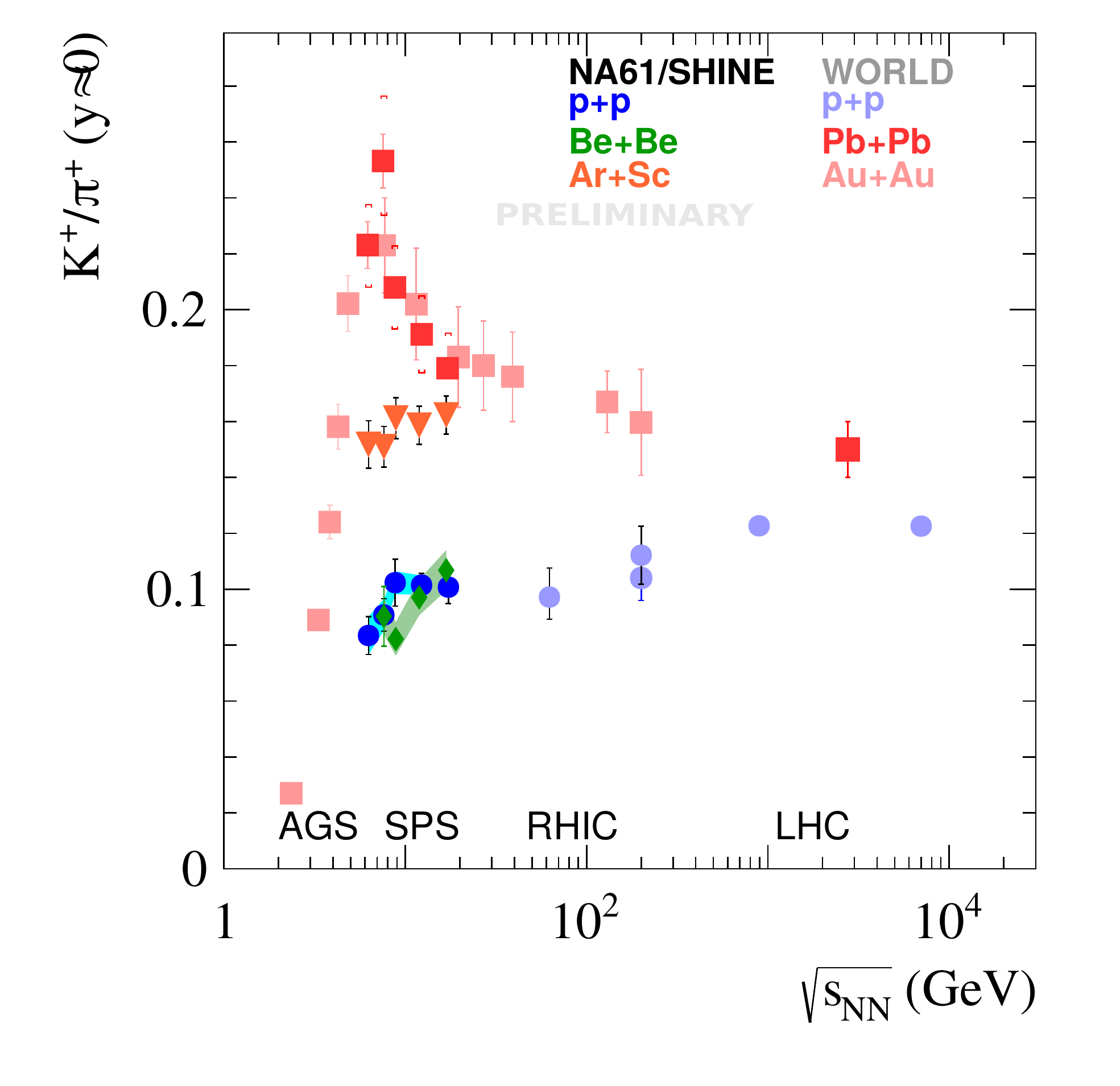}			
			\end{minipage}
			\begin{minipage}{0.34\linewidth}
			\includegraphics[width=\textwidth]{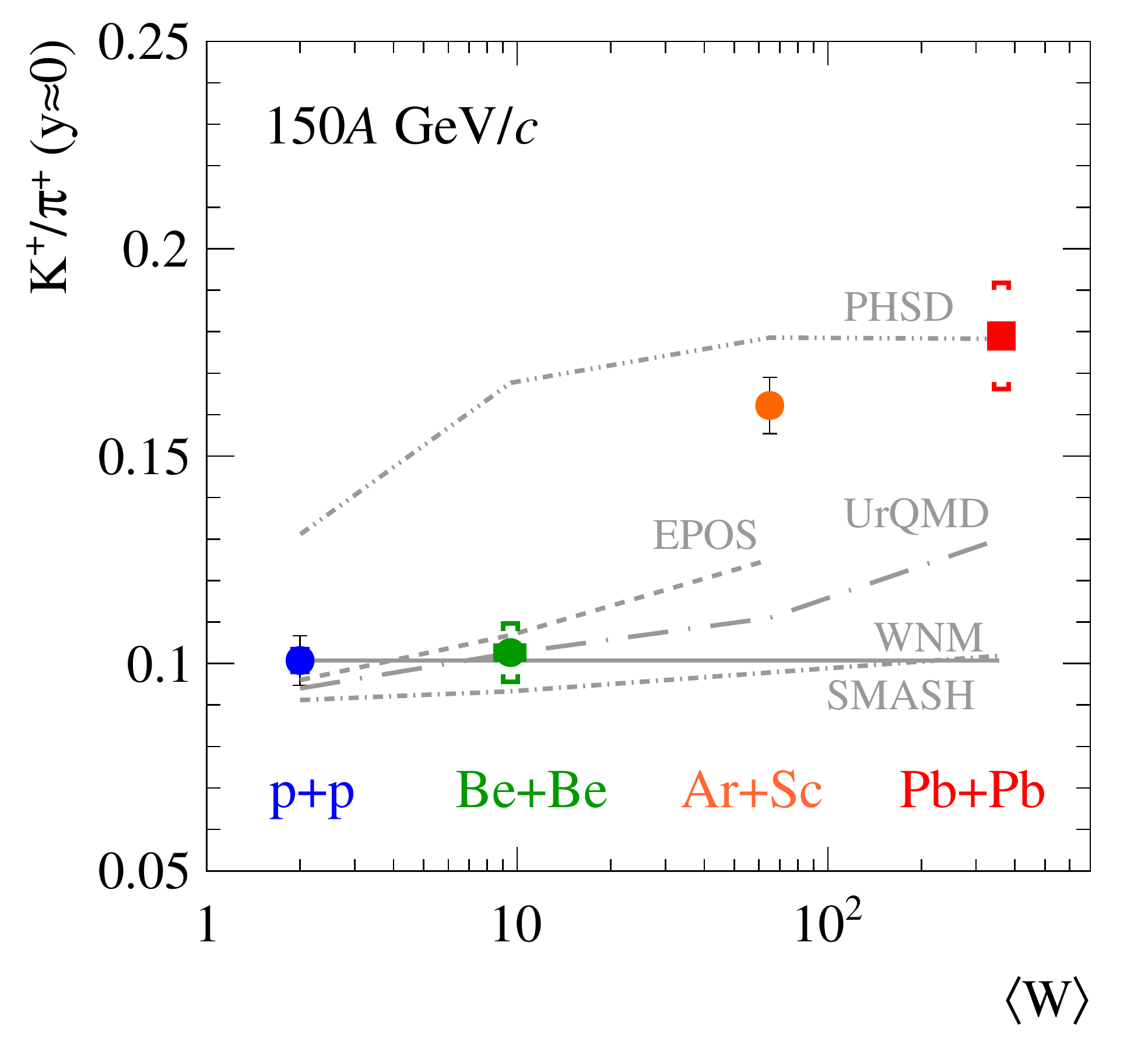}
				\\
			\includegraphics[width=\textwidth]{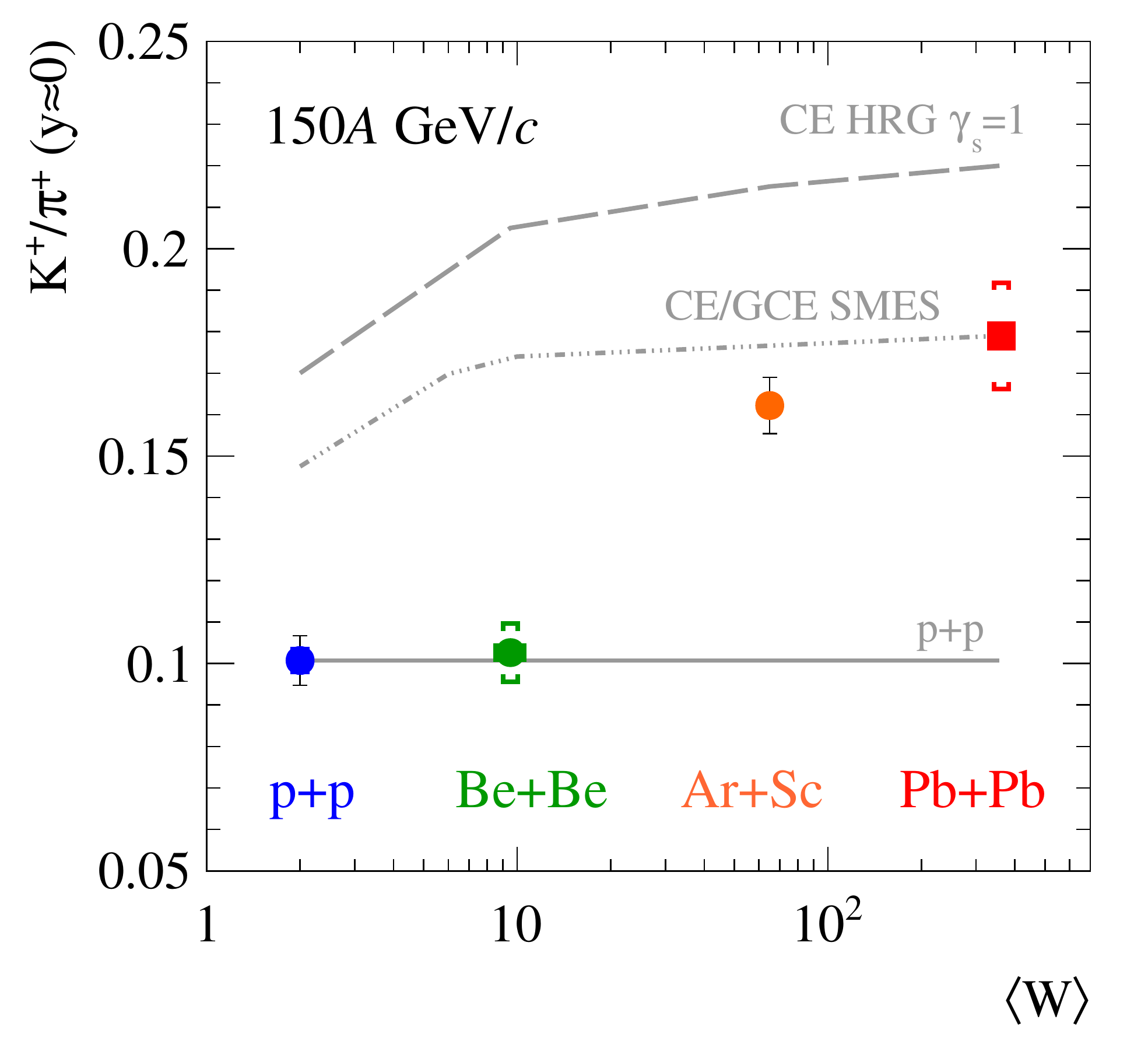}	
			\end{minipage}
			\label{fig:horn}
			\caption{System-size and energy dependence of $K^{+}/\pi^{+}$ ratio compared with other experiments (left panel) as well as string (right top panel) and statistical (right bottom panel) models. For details see text. }
	\end{figure}

\begin{figure}
			\centering	
			\begin{overpic}[width=0.4\textwidth]{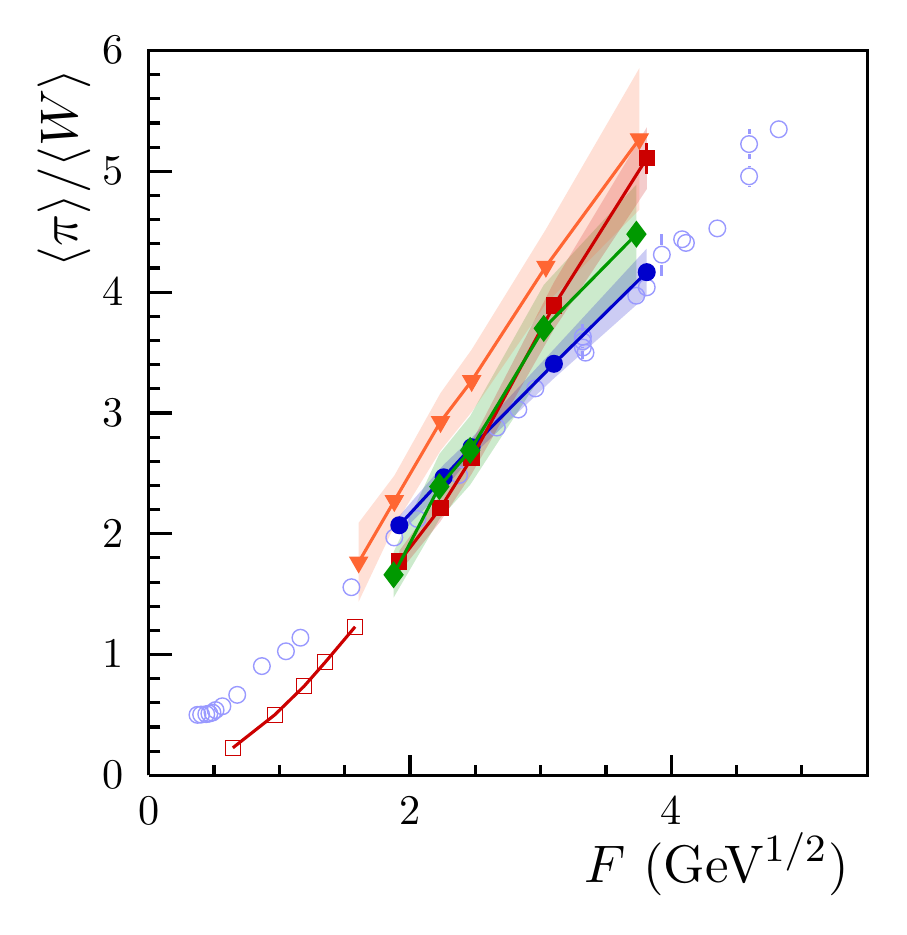}
			\put(46,20){\includegraphics[width=0.15\textwidth]{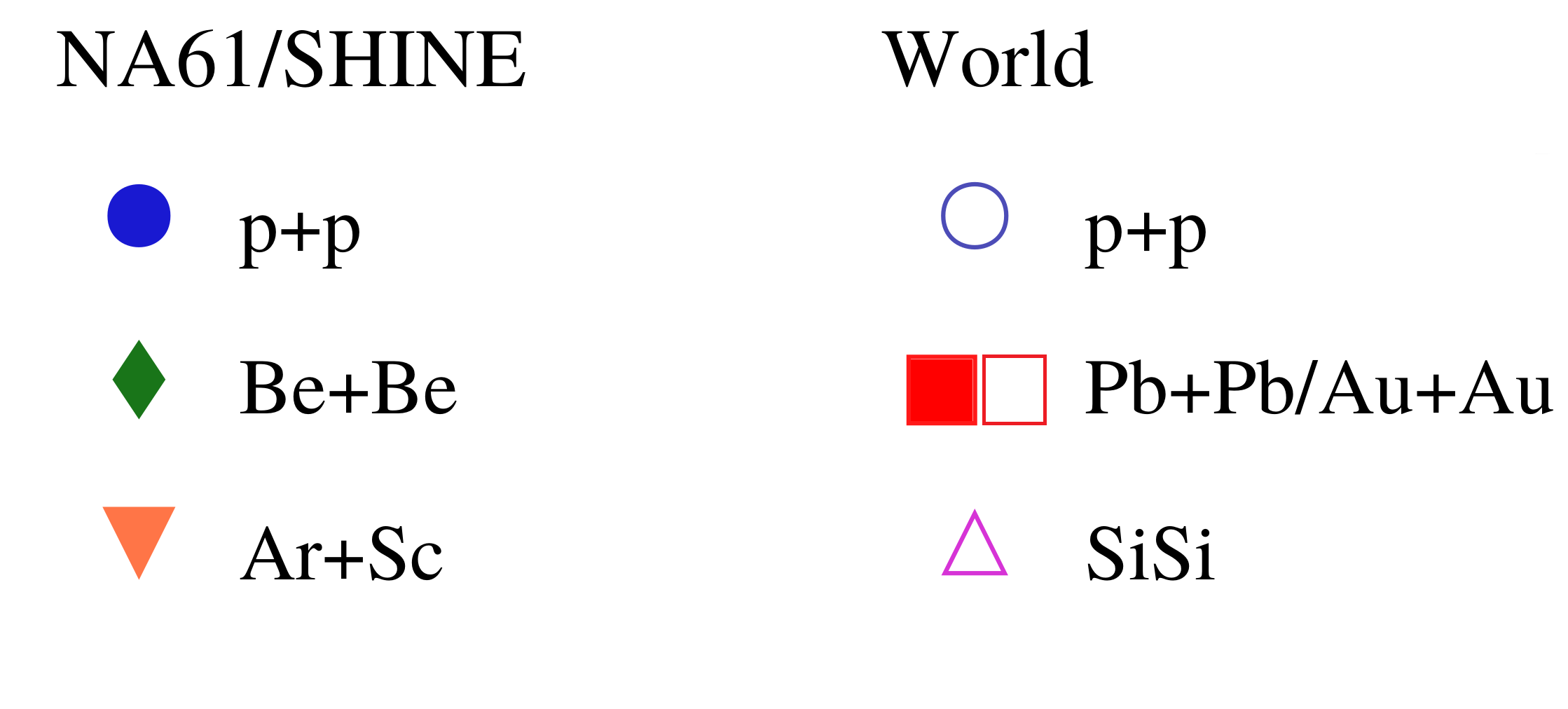}}
			\end{overpic}
						
			\caption{Mean pion multiplicity to the mean number of wounded nucleons versus the Fermi
energy measure.}
			\label{fig:kink}
	\end{figure}

Figure~\ref{fig:horn} (left) presents energy and system-size dependence of $K^{+}/\pi^{+}$ ratio at mid-rapidity obtained by NA61/SHINE in p+p~\cite{pp}, Be+Be~\cite{BeBe} and Ar+Sc (preliminary); and NA49 in Pb+Pb~\cite{NA49PbPb}. The so-called horn in the $K^{+}/\pi^{+}$ ratio was predicted within SMES as one of the signatures of the OD. The data were compared with few dynamical models (see right top panel of Fig.~\ref{fig:horn}). Those without phase transition (EPOS~\cite{Pierog:2009zt, EPOSWeb}, UrQMD~\cite{UrQMD} and SMASH~\cite{SMASH}) agree with the results from small systems (p+p and Be+Be), while do not describe the results from heavier systems (Ar+Sc and Pb+Pb). In contrast, the model with phase transition (PHSD~\cite{PHSD}), follows the trend observed at the heaviest system (Pb+Pb), but overestimates the ratio for smaller systems. Both tested statistical models: renormalised SMES~\cite{SMES2} (with the phase transition) and HRG~\cite{HRG} (without it) overestimate the $K^{+}/\pi^{+}$ ratio especially in small systems (see right bottom panel of Fig.~\ref{fig:horn}).
The energy and system-size dependence of another signature - the so-called kink, is shown in Fig.~\ref{fig:kink}. It presents the dependence of the ratio of the mean number of pions to the mean number of wounded nucleons $\langle\pi\rangle/\langle W\rangle$ versus the Fermi energy measure $F=\big[ \frac{\sqrt{s_{NN}-2m_{N}}}{\sqrt{s_{NN}}}\big]^{1/4}\approx \sqrt[4]{s_{NN}}$. At low energies $\langle\pi\rangle/\langle W\rangle$ for Ar+Sc reactions equals that for N+N\footnote{For p+p interactions the figure shows isospin symmetrized values~\cite{iso_pp} marked as N+N.}. At high SPS energies it becomes consistent with central Pb+Pb interactions. The behavior of Ar+Sc stands in contradiction to Be+Be measurements, which are close to the Pb+Pb results except for the top SPS beam energy. 

	Spatial asymmetry of the initial energy density in the overlapping region of the colliding relativistic nuclei is converted to the asymmetry of momentum distribution of particles in the final state. The resulting asymmetry carries information about the transport properties of the created quark and gluons system. Asymmetry is usually quantified with $v_n$ coefficients in a Fourier decomposition of the azimuthal distribution of produced particles relative to the reaction plane. The Projectile Spectator Detector allows NA61/SHINE to estimate the reaction plane in a unique way (for details see Refs.~\cite{Golosov:2019sdu,EKashirin}). 

\begin{figure}
			\centering	
			\includegraphics[width=0.32\textwidth]{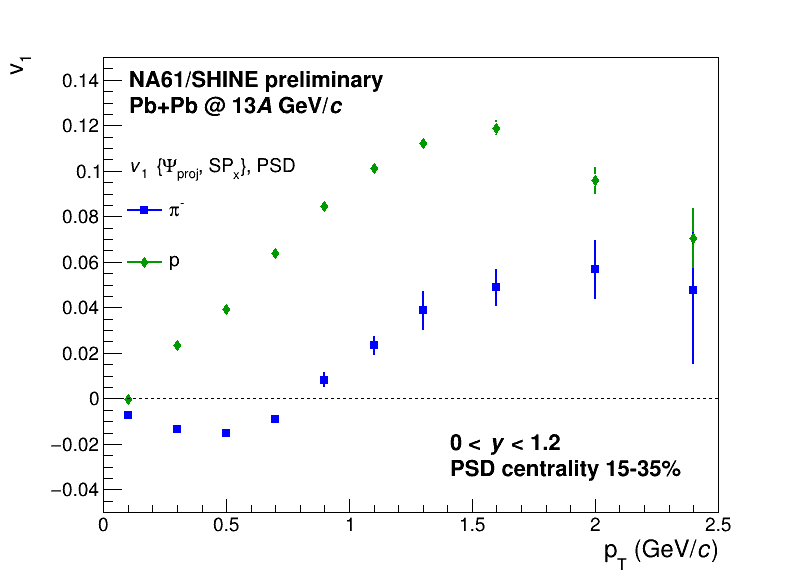}
			\includegraphics[width=0.305\textwidth]{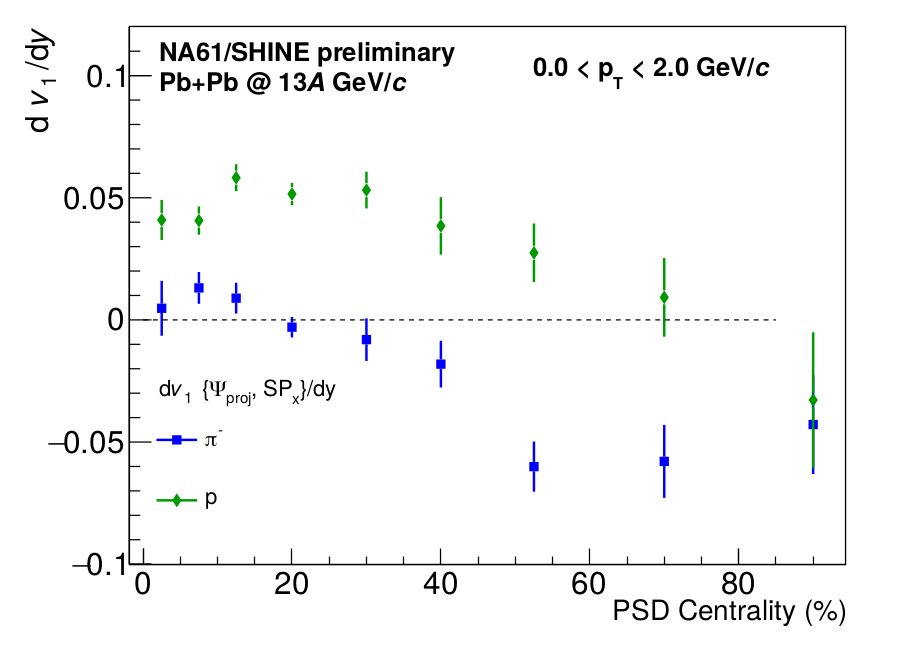}
			\includegraphics[width=0.32\textwidth]{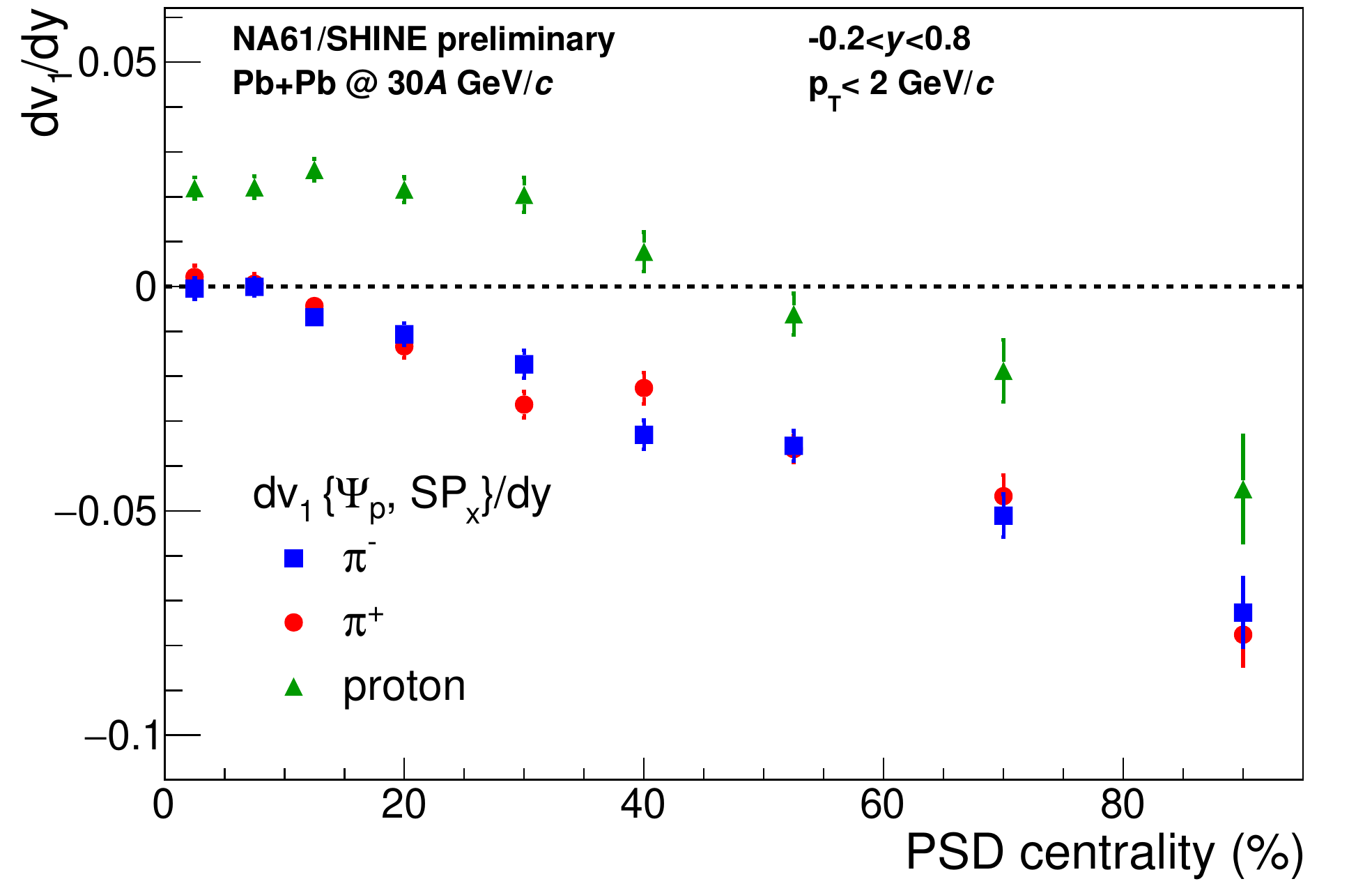}	
			\vspace{-0.15in}		
			\caption{Negatively charged pion and proton directed flow $v_{1}(p_{T})$ and $dv_{1}/dy$ for different centrality classes in Pb+Pb collisions.}
			\label{fig:flow}
	\end{figure}

The slope of proton directed flow at mid-rapidity, $dv_1/dy$, at considered energy range it is expected to change its sign~\cite{STARv,STARv2,Wu:2018qih}. Directed flow of $\pi^{-}$ and $p$ as well as $dv_1/dy$ (centrality dependence) for Pb+Pb collisions at 13$A$ and 30$A$ GeV/$c$ is presented in Fig.~\ref{fig:flow}. Shapes of $v_{1}(p_{T})$ for $p$ and $\pi^{-}$ (Fig.~\ref{fig:flow}, left) are different: $v_1(p_T)$ of $p$ is positive in the entire $p_T$ range while directed flow of $\pi^{-}$ starts with negative values and than changes sign (Fig.~\ref{fig:flow}, center). There is also a clear difference of the $v_{1}$ slope between 13$A$ and 30$A$ GeV/$c$ (Fig.~\ref{fig:flow}, center and right).
\section{Search for the critical point}

\begin{figure}
			\centering	
			\includegraphics[width=0.32\textwidth]{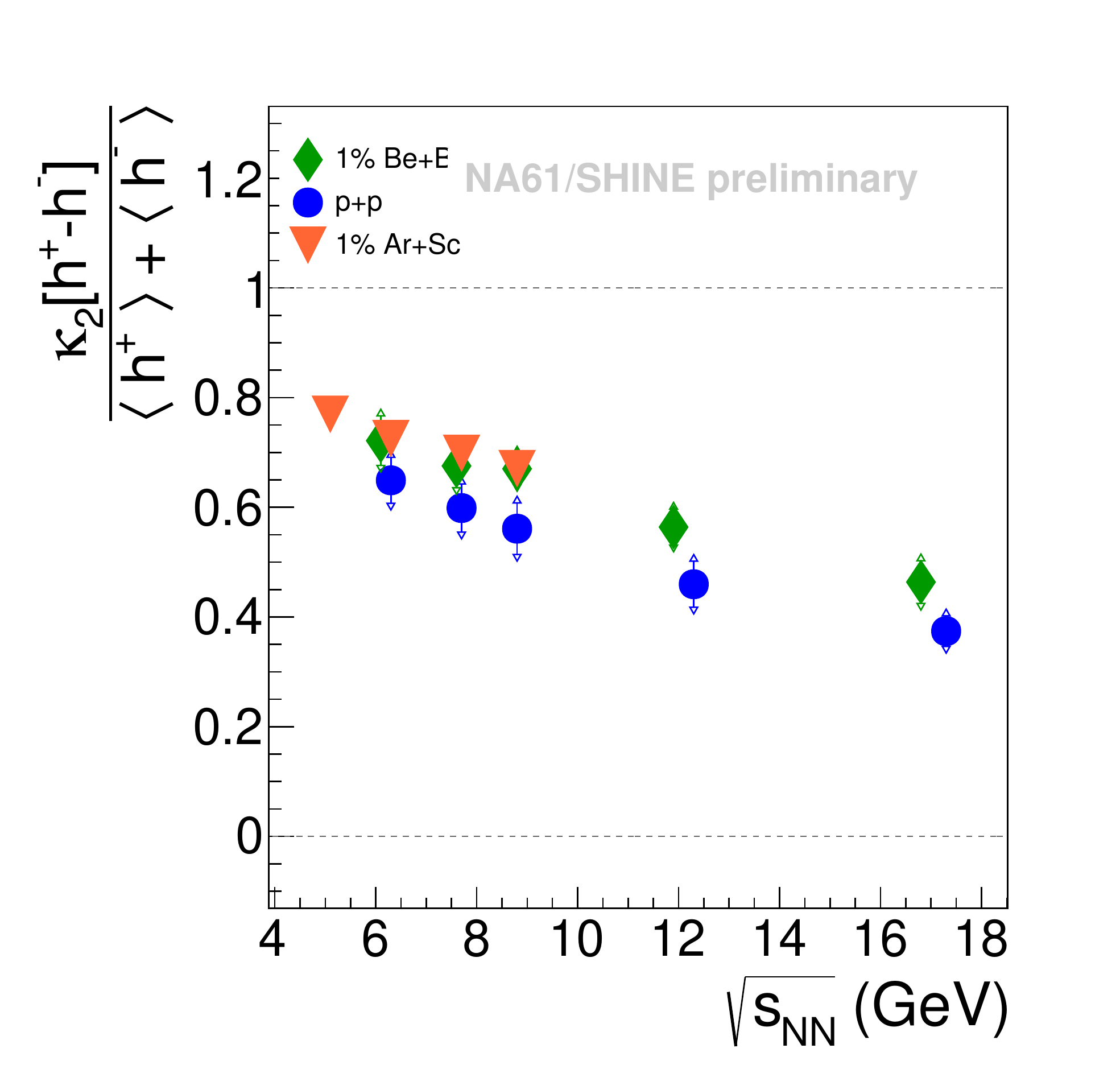}
			\includegraphics[width=0.32\textwidth]{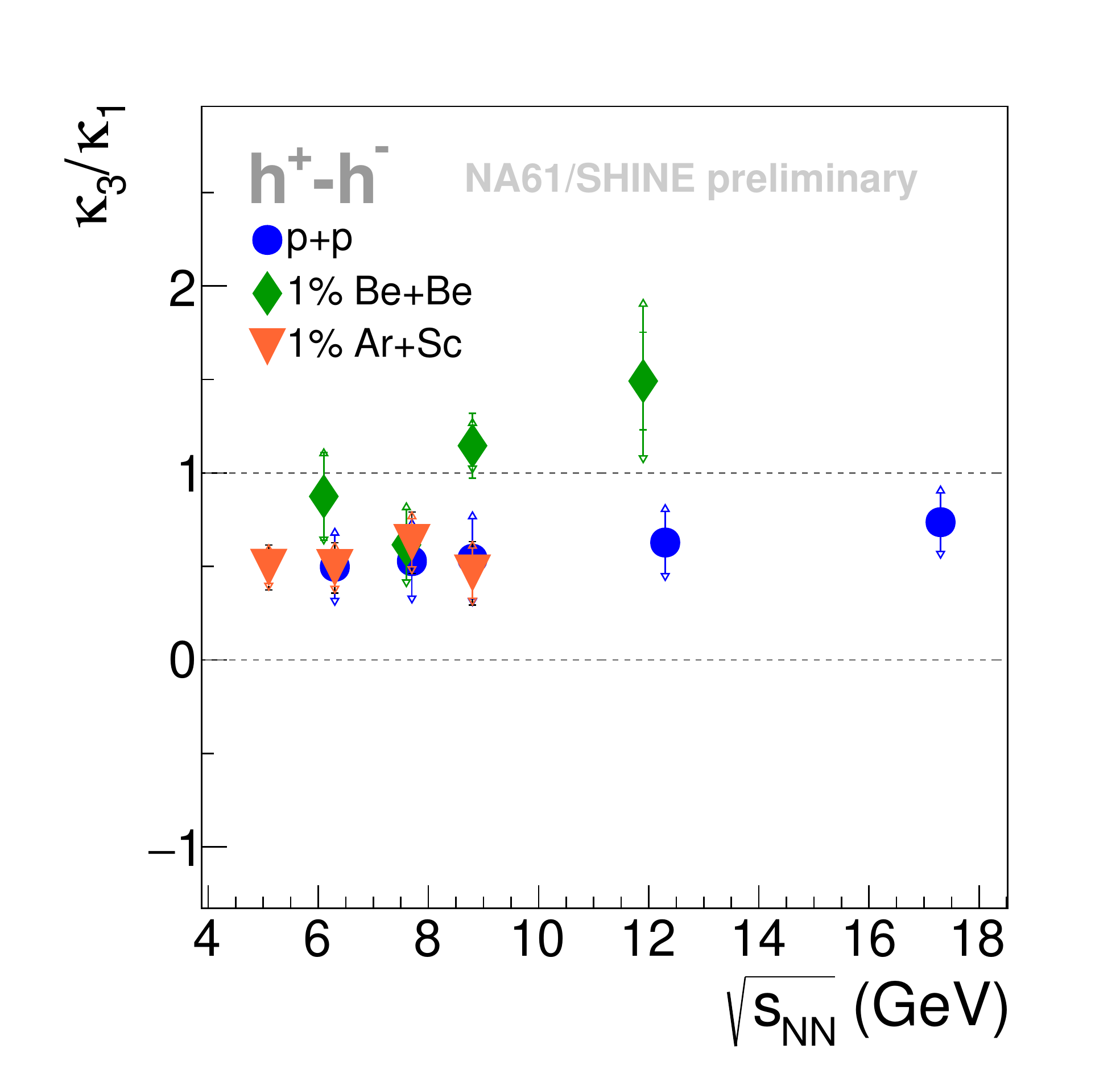}
			\includegraphics[width=0.32\textwidth]{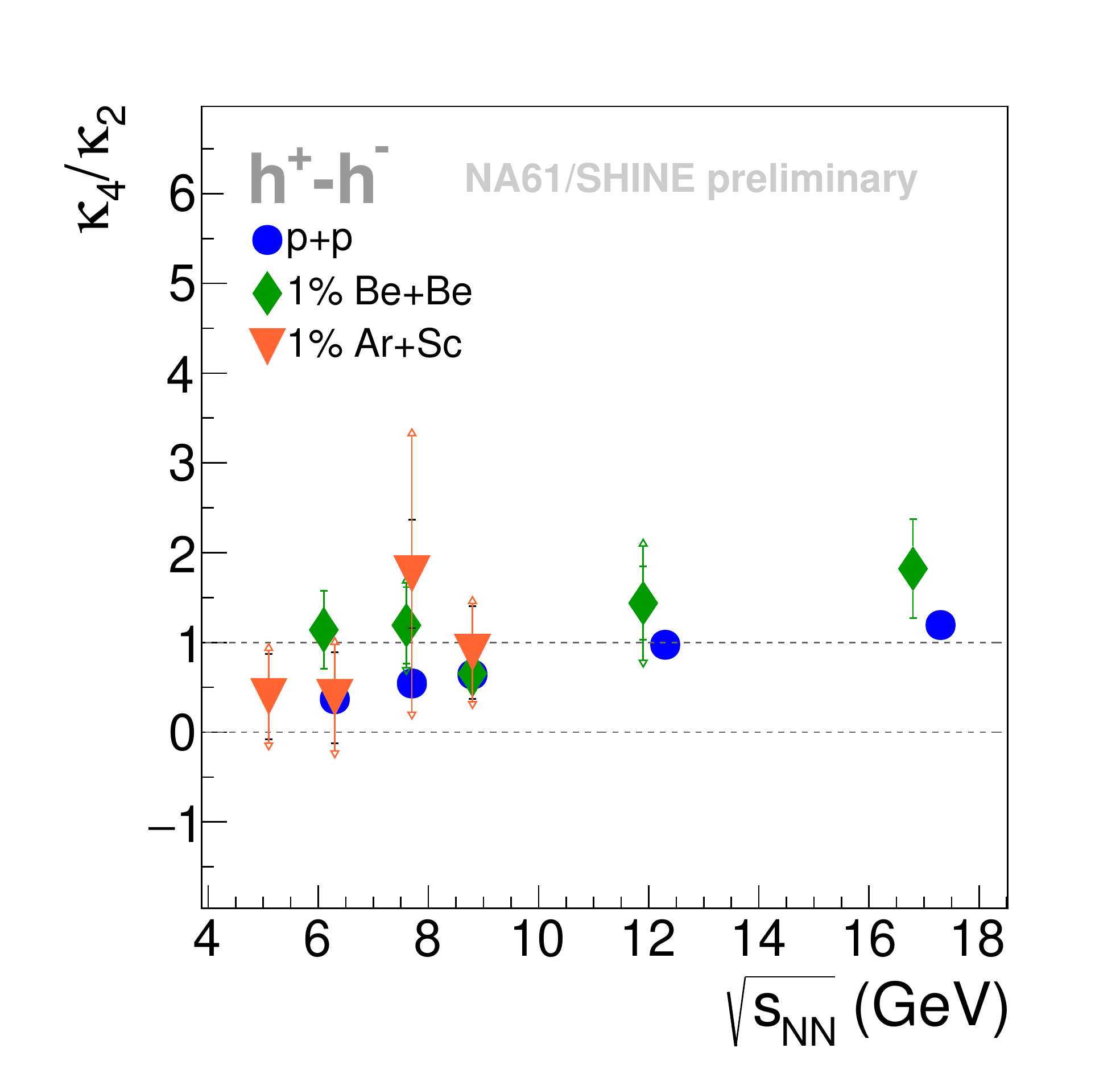}
			\vspace{-0.15in}
			\caption{System size and energy dependence of $\kappa_{2}/(\kappa_{1}[h^{+}]-\kappa_{1}[h^{-}])$, $\kappa_{3}/\kappa_{1}[h^{+}-h^{-}]$ and $\kappa_{4}/\kappa_{2}[h^{+}-h^{-}]$. Statistical uncertainty was obtained with the bootstrap method and it is indicated as a dashed black bar. Systematic uncertainty/bias: p+p - corrected data with estimate on systematic uncertainty; Be+Be/Ar+Sc - uncorrected data with estimate of systematic bias. Systematic uncertainty/bias is indicated with arrows.}
			\label{fig:net}
	\end{figure}
	
	The expected signal of a critical point (CP) is a non-monotonic dependence of various fluctuation/correlation measures in NA61/SHINE energy -- system size scan. Fluctuations of conserved charges (electric, strangeness or baryon number) are of special interest~\cite{Stephanov_overview, Asakawa:2015ybt,Stephanov:1998dy}.

To compare fluctuations in systems of different sizes, one should use quantities insensitive to system volume, i.e. intensive quantities. They are constructed by division of cumulants $\kappa_{i}$ of the measured multiplicity distribution (up to fourth order), where $i$ is the order of the cumulant. For second, third and fourth order cumulants intensive quantities are defined as: $\kappa_{2}/\kappa_{1}$, $\kappa_{3}/\kappa_{2}$ and $\kappa_{4}/\kappa_{2}$. Their reference values for no fluctuations are 0 and for independent particle production are 1. In case of net-charge, cumulant ratios are redefined to $\kappa_{2}/(\kappa_{1}[h^{+}]-\kappa_{1}[h^{-}])$, $\kappa_{3}/\kappa_{1}[h^{+}-h^{-}]$ and $\kappa_{4}/\kappa_{2}[h^{+}-h^{-}]$ in order to keep the same references. 

Figure~\ref{fig:net} shows the system size and energy dependence of second, third and fourth order cumulant ratio of net-electric charge in p+p as well as central Be+Be and Ar+Sc interactions. So far, there is no clear difference between systems for higher order moments. More detailed studies are needed.

\begin{figure}
			\centering	
			\includegraphics[width=0.32\textwidth]{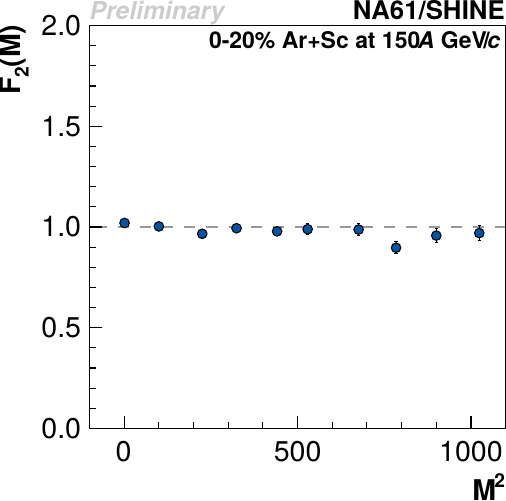}
			\includegraphics[width=0.32\textwidth]{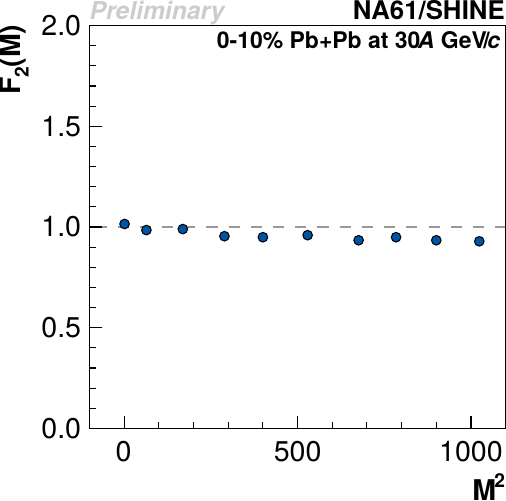}
			\vspace{-0.15in}
			\caption{Preliminary results on $F_2(M)$ of mid-rapidity protons measured in 0-20$\%$ most central Ar+Sc collisions at 150$A$ GeV/$c$ (left) and 0-10$\%$ most central Pb+Pb collisions (right).}
			\label{fig:F2}
	\end{figure}
Another, possible tool for search of CP is a proton intermittency. In the proximity of CP a local power-law fluctuations of the baryon density should appear which can be searched for by studying second factorial moments, 
$F_{2}(\delta)=\frac{\langle\frac{1}{M}\sum_{i=1}^{M}n_{i}(n_{i}-1)\rangle}{\langle\frac{1}{M}\sum_{i=1}^{M}n_{i}\rangle^{2}}$
with the cell size or, equivalently, with the number of cells in ($p_x , p_y$ ) space of protons at
mid-rapidity~\cite{Bialas:1985jb, Turko:1989dc,Diakonos:2006zz}. NA61/SHINE measures $F_2(M)$ using using statistically independent
points and cumulative variables. Preliminary results on $F_2(M)$ of mid-rapidity protons measured in 0-20$\%$ most central Ar+Sc collisions at 150$A$ GeV/$c$ and 0-10$\%$ most central Pb+Pb
collisions are presented in Fig.~\ref{fig:F2} in left and right panels, respectively. The intermittency index
$\phi_2$ for a system freezing out at the QCD critical endpoint is expected to be $\phi_{2} = 5/6$ assuming
that the latter belongs to the 3-D Ising universality class. Measured $F_2(M)$ of protons for Ar+Sc at 150$A$
GeV/$c$ and Pb+Pb at 30$A$ GeV/$c$ show no indication for power-law increase with a bin size which could indicate CP.

\section{Summary}
Presented experimental results show unexpected system-size dependence of $K^{+}/\pi^{+}$ ratio at mid-rapidity which is not described by studied models. The results on energy dependence of $\langle\pi\rangle/\langle W\rangle$ ratio suggests an increase of effective number of degrees of freedom in central Ar+Sc collisions at top SPS energies. Presented results do not show any indications of the CP in the presented analysis. Clear separation of small (p+p, Be+Be) and large (Ar+Sc, Pb+Pb) systems for spectra results does not remain for fluctuations of net-electric charge described by higher-order cumulants of the distribution.

\textbf{Acknowledgments:} This work was supported by WUT-IDUB and the National Science Centre, Poland under grant no. 2016/21/D/ST2/01983.

\end{document}